# Computing Boolean Functions: Exact Quantum Query Algorithms and Low Degree Polynomials

Alina Dubrovska, Taisija Mischenko-Slatenkova

Department of Computer Science, University of Latvia,
Raina bulv. 29, LV-1459, Riga, Latvia
alina.dubrovska@gmail.com, stajka@inbox.lv

**Abstract.** In this paper we study the complexity of quantum query algorithms computing the value of Boolean function and its relation to the degree of algebraic polynomial representing this function. We pay special attention to Boolean functions with quantum query algorithm complexity lower than the deterministic one. Relation between the degree of representing polynomial and potentially possible quantum algorithm complexity has been already described; unfortunately, there are few examples of quantum algorithms to illustrate theoretical evaluation of the complexity. Work in this direction was aimed (1) to construct effective quantum query algorithms for computing Boolean functions, (2) to design methods for Boolean function construction with a large gap between deterministic complexity and degree of representing polynomial. In this paper we present our results in both directions.

## 1  Introduction

Let $f(x_1, x_2, ..., x_n): \{0,1\}^n \to \{0,1\}$ be a Boolean function. We study the query model, where the input $(x_1, x_2, ..., x_n)$ is contained in a black box and can be accessed by asking questions about the values of $x_i$. Goal here is to compute the value of function. The complexity of a query algorithm is measured in number of questions that it asks. The classical version of this model is known as *decision trees* (for details see [1]). Quantum query algorithms can solve certain problems faster than classical algorithms. Best known exact quantum algorithm is for *PARITY* function with $n/2$ questions vs. $n$ questions required by classical algorithm. In our algorithms we are trying to use quantum parallelism feature to obtain best results.

Every Boolean function can be represented by an algebraic polynomial, which is unique. It has been proved, that degree of such polynomial is related to query algorithms complexity. For quantum computation important are functions with low polynomial degree.

## 2 Definitions

We use $\oplus$ to denote XOR (exclusive OR). We use $\bar{f}$ for the function $1 - f$.

### 2.1 Quantum computing

We use the basic model of quantum computation (for details see textbooks [2], [3]).

An *n*-dimensional quantum state is a vector $|\psi\rangle \in C^n$ of norm 1. Let $|0\rangle, |1\rangle, \ldots, |n-1\rangle$ be an orthonormal basis for $C^n$. Then, any state can be expressed as $|\psi\rangle = \sum_{i=0}^{n-1} a_i |i\rangle$ for some $a_0 \in C, a_1 \in C, \ldots, a_{n-1} \in C$. Since the norm of $|\psi\rangle$ is 1, we have $\sum_{i=0}^{n-1} |a_i|^2 = 1$. States $|0\rangle, |1\rangle, \ldots, |n-1\rangle$ are called *basic states*. Any state of the form $\sum_{i=0}^{n-1} a_i |i\rangle$ is called a *superposition* of $|0\rangle, |1\rangle, \ldots, |n-1\rangle$. The coefficient $a_i$ is called *amplitude* of $|i\rangle$.

State of a system can be changed using *unitary transformations*. Unitary transformation $U$ is a linear transformation on $C^n$ that maps vectors of unit norm to vectors of unit norm.

The simplest case of quantum measurement is used in our model. It is the full measurement in the computation basis. Performing this measurement on a state $|\psi\rangle = a_1|0\rangle + \ldots a_k|k\rangle$ gives the outcome $i$ with probability $|a_i|^2$.

### 2.2 Query model

We consider computing Boolean functions in the quantum query model (for details see survey [4]).

A quantum computation with $T$ queries is just a sequence of unitary transformations

$$U_0 \to Q \to U_1 \to Q \to \ldots \to U_{T-1} \to Q \to U_T$$

$U_i$'s can be arbitrary unitary transformations that do not depend on the input bits $x_1, x_2, \ldots, x_n$. $Q$'s are query transformations. The computation starts with a state $|0\rangle$. Then, we apply $U_0, Q, \ldots, Q, U_T$ and measure the final state.

We use extension of sign query in our model. To specify each question we have to assign a number of queried variable to each amplitude. A query will change sign of amplitude to opposite if value of assigned variable is 1 and leave as it is otherwise.

Each amplitude of final quantum state corresponds to algorithm output. We assign a value of a function to each output. The result of running algorithm on input *X* is *j* with probability which equals the sum of squares of all amplitudes, which corresponds to outputs with value *j*.

Very convenient way of query algorithm representation is a graphical picture, where each horizontal line corresponds to amplitude.

### 2.3 Query Complexity and Polynomials

Let $D(f)$ be the deterministic decision tree complexity and $Q_E(f)$ be the exact quantum query algorithm complexity (see [1] for definitions).

Sensitivity of *f* on input $(x_1, x_2, \ldots, x_n)$ is the number of variables $x_i$ with property that $f(x_1,\ldots,x_i,\ldots,x_n) \neq f(x_1,\ldots,1-x_i,\ldots,x_n)$. Sensitivity of *f* is maximum sensitivity of all possible inputs. It has been proved, that $s(f) \leq D(f)$. In particular, if $s(f)$ is equal to the number of variables *n*, then $D(f)=n$.

The degree of the representing polynomial is called the degree of the Boolean function and is denoted as *deg(f)*. We have $D(f) \geq \deg(f)$ [5] and $Q_E(f) \geq \frac{\deg(f)}{2}$ [1].

Hamming weight of the input *x* is denoted as |*x*| and is equal to the number of variables $x_i=1$.

## 3 Main Results

We introduce our results consisting of two parts. First, we present new exact quantum algorithms, the best of which provides the same gap from deterministic complexity as famous *PARITY* algorithm. Then we describe low-degree polynomials with relation between the number of variables and the degree greater than 2 in the best case.

### 3.1 Exact Quantum Algorithms

In this section we show exact quantum algorithms that can be used to compute large sets of functions faster than deterministic algorithms.

**Exact Quantum Algorithm with $\frac{2}{3}D(f)$ Queries**

First, let's consider function of 3 variables $F_3(x_1, x_2, x_3) = \neg(x_1 \oplus x_2) \wedge (x_1 \oplus x_3)$.

**Theorem 1** $D(F_3) = 3$.

**Proof.** Provided by sensitivity on any $x$ such that $F_3(x) = 1$ ($x = 001, x = 110$). □

Function $F_3$ can be represented by a polynomial of degree 2:
$$p(x) = \frac{1}{2}(x_1 x_2 + \bar{x}_1 \bar{x}_2) + \frac{1}{2}(1 - (x_1 x_3 + \bar{x}_1 \bar{x}_3)) - \frac{1}{2}(x_2 x_3 + \bar{x}_2 \bar{x}_3)$$

This gives us a hope that exact quantum query algorithm exists, which is more efficient, than deterministic. We confirm this hypothesis proposing definite algorithm.

**Theorem 2** *Exact quantum algorithm $A_1$ exists, which computes $F_3$ with 2 queries.*

**Proof.** Algorithm is presented in Figure 1.

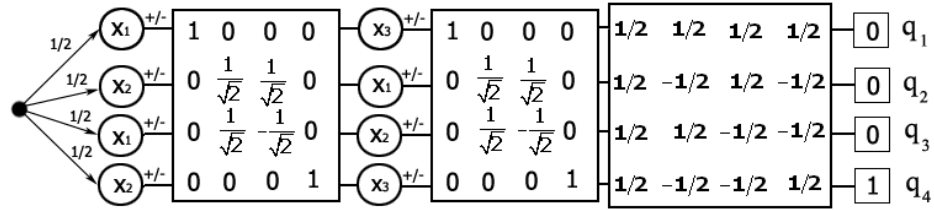

**Fig. 1.** Quantum query algorithm $A_1$

Two qubits are used for computation, so the total number of amplitudes is 4. We start with state $|\vec{0}\rangle$, apply unitary transformations, then measure the final state and always see correct result with probability 1 for any input.

Following unitary transformations are used in computation:

$$U_0 = \frac{1}{2}\begin{pmatrix} 1 & 1 & 1 & 1 \\ 1 & -1 & 1 & -1 \\ 1 & 1 & -1 & -1 \\ 1 & -1 & -1 & 1 \end{pmatrix} \quad Q_1 = \begin{pmatrix} (-1)^{x_1} & 0 & 0 & 0 \\ 0 & (-1)^{x_2} & 0 & 0 \\ 0 & 0 & (-1)^{x_1} & 0 \\ 0 & 0 & 0 & (-1)^{x_2} \end{pmatrix}$$

$$U_1 = \begin{pmatrix} 1 & 0 & 0 & 0 \\ 0 & \frac{1}{\sqrt{2}} & \frac{1}{\sqrt{2}} & 0 \\ 0 & \frac{1}{\sqrt{2}} & -\frac{1}{\sqrt{2}} & 0 \\ 0 & 0 & 0 & 1 \end{pmatrix} \quad Q_2 = \begin{pmatrix} (-1)^{x_3} & 0 & 0 & 0 \\ 0 & (-1)^{x_1} & 0 & 0 \\ 0 & 0 & (-1)^{x_2} & 0 \\ 0 & 0 & 0 & (-1)^{x_3} \end{pmatrix}$$

Computation process is specified by following sequence:
$$|0\rangle \rightarrow U_0 \rightarrow Q_1 \rightarrow U_1 \rightarrow Q_2 \rightarrow U_1 U_0 \rightarrow [M]$$

As example we show complete computation process for input $x=011$.

$$(1,0,0,0) \xrightarrow{U0} \left(\frac{1}{2},\frac{1}{2},\frac{1}{2},\frac{1}{2}\right) \xrightarrow{Q1} \left(\frac{1}{2},-\frac{1}{2},\frac{1}{2},-\frac{1}{2}\right) \xrightarrow{U1} \left(\frac{1}{2},0,-\frac{1}{\sqrt{2}},-\frac{1}{2}\right) \xrightarrow{Q2}$$

$$\xrightarrow{Q2} \left(-\frac{1}{2},0,\frac{1}{\sqrt{2}},\frac{1}{2}\right) \xrightarrow{U1} \left(-\frac{1}{2},\frac{1}{2},-\frac{1}{2},\frac{1}{2}\right) \xrightarrow{U0} (0,-1,0,0) \xRightarrow{[M]} F_3(011) = 0$$

□

**Theorem 3** *Every Boolean function* $f(x_1, x_2, x_3)$ *with a property* $f(X) = f(\bar{X})$ *can be computed by exact quantum query algorithm with 2 queries.*

**Proof.** We examine the process of how $A_1$ compute $F_3$ for different inputs. We are interested in final distribution of amplitudes before the measurement. Figure 2 provides complete picture and some specific symmetry can be noticed.

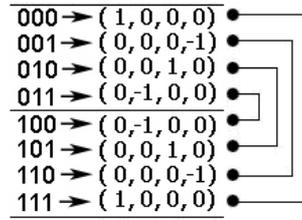

**Fig. 2.** Symmetry in final distribution of amplitudes for different inputs for algorithm $A_1$

We can assign values to outputs at one's own choice. Algorithm $A_1$ can be modified for each function of concerned set by choosing assignment in suitable manner. To obtain the result we are searching for, we have to use the following assignment rule:

$$(f(000) = a) \leftrightarrow (q_1 \equiv a) \quad (f(010) = a) \leftrightarrow (q_3 \equiv a)$$
$$(f(001) = a) \leftrightarrow (q_4 \equiv a) \quad (f(011) = a) \leftrightarrow (q_2 \equiv a) \quad, a \in \{0,1\}$$

□

From all functions with property $f(x) = f(\bar{x})$ we select only those with deterministic complexity 3 and denote this set with $S3$.

**Table 1.** Set $S3$ of functions with a gap of 2 vs. 3 between deterministic and quantum query complexity

| X | $f_1$ | $f_2$ | $f_3$ | $f_4$ | $f_5$ | $f_6$ | $f_7$ | $f_8$ |
|---|---|---|---|---|---|---|---|---|
| 000 | 1 | 0 | 0 | 0 | 1 | 1 | 1 | 0 |
| 001 | 0 | 1 | 0 | 0 | 1 | 1 | 0 | 1 |
| 010 | 0 | 0 | 1 | 0 | 1 | 0 | 1 | 1 |
| 011 | 0 | 0 | 0 | 1 | 0 | 1 | 1 | 1 |
| 100 | 0 | 0 | 0 | 1 | 0 | 1 | 1 | 1 |
| 101 | 0 | 0 | 1 | 0 | 1 | 0 | 1 | 1 |
| 110 | 0 | 1 | 0 | 0 | 1 | 1 | 0 | 1 |
| 111 | 1 | 0 | 0 | 0 | 1 | 1 | 1 | 0 |

Next we will switch from 2 vs. 3 gap to $2n$ vs. $3n$.

**Theorem 4** [7] *Let Q be an exact quantum query algorithm, which computes Boolean function $f_1(x_1,...,x_m)$ with $k_1$ queries. Corresponding deterministic algorithm requires $k_2$ queries ($k_2>k_1$). Let D be a deterministic query algorithm, which computes Boolean function $f_2(x_1,...,x_n)$ with n queries. Then exact quantum query algorithm Q' exists, which computes function $f_2(f_1(x_1,...,x_m),..,f_1(x_{(n-1)m+1},...,x_{nm}))$ with $k_1n$ queries and corresponding deterministic query algorithm have $k_2n$ queries.*

**Theorem 5** *Each function from set S3 can be used as a base for constructing composite Boolean function f and this function will have exact quantum query algorithm Q with complexity $Q_E(Q) = \frac{2}{3}D(f)$.*

**Proof.** We can generalize each function $f_i$ from set *S*3 in a way described in Theorem 4. We can take any Boolean function $h$ of $n$ variables with $D(h) = n$ and construct new function $f = h(f_i, f_i,..., f_i)$. Deterministic complexity of such function is $D(f) = 3n$ and exact quantum query algorithm Q exists with $Q_E(Q) = 2n$. From here follows the relation $Q_E(Q) = \frac{2}{3}D(f)$. □

# Exact quantum algorithm with $\frac{D(f)}{2}$ queries

Based on results described in previous section we have decided to take a look to Boolean functions of 4 variables with the same property $f(X) = f(\bar{X})$.

Let's consider the function: $G_4(x_1,x_2,x_3,x_4) = (x_1 \oplus x_2) \wedge (x_3 \oplus x_4)$.

**Theorem 6** $D(G_4) = 4$.

**Proof.** Provided by sensitivity on any $x$ such that $G_4(x) = 1$ (for example $x$=0101). □

We have succeeded in attempt to construct exact quantum query algorithm with less queries than deterministic algorithm.

**Theorem 7** *Exact quantum algorithm $A_2$ exists, which computes $G_4$ with 2 queries.*

**Proof.** Algorithm is presented in Figure 3.

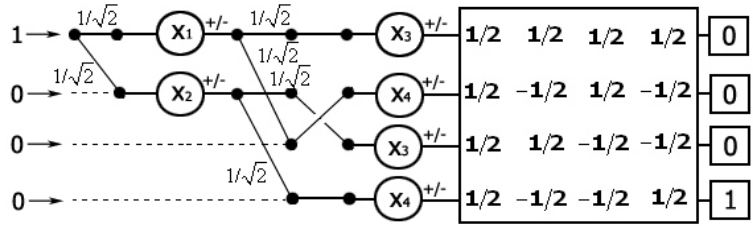

**Fig. 3.** Quantum query algorithm $A_2$

Result of computation equals $G_4(x)$ for any input $x$ with probability 1.

□

Again, we can use the same approach of changing the assignment of values to outputs from the proof of Theorem 3 to modify $A_2$ to compute the whole set of functions. We select only functions with deterministic complexity 4 and get the set $S4$.

**Table 2.** Set $S4$ of functions with a gap of 2 vs. 4 between deterministic and quantum query complexities

| X | $g_1$ | $g_2$ | $g_3$ | $g_4$ | $g_5$ | $g_6$ | $g_7$ | $g_8$ | X | $g_1$ | $g_2$ | $g_3$ | $g_4$ | $g_5$ | $g_6$ | $g_7$ | $g_8$ |
|---|---|---|---|---|---|---|---|---|---|---|---|---|---|---|---|---|---|
| 0000 | 0 | 0 | 0 | 1 | 1 | 1 | 1 | 0 | 1000 | 0 | 0 | 1 | 0 | 1 | 1 | 0 | 1 |
| 0001 | 0 | 1 | 0 | 0 | 1 | 0 | 1 | 1 | 1001 | 1 | 0 | 0 | 0 | 0 | 1 | 1 | 1 |
| 0010 | 0 | 1 | 0 | 0 | 1 | 0 | 1 | 1 | 1010 | 1 | 0 | 0 | 0 | 0 | 1 | 1 | 1 |
| 0011 | 0 | 0 | 0 | 1 | 1 | 1 | 1 | 0 | 1011 | 0 | 0 | 1 | 0 | 1 | 1 | 0 | 1 |
| 0100 | 0 | 0 | 1 | 0 | 1 | 1 | 0 | 1 | 1100 | 0 | 0 | 0 | 1 | 1 | 1 | 1 | 0 |
| 0101 | 1 | 0 | 0 | 0 | 0 | 1 | 1 | 1 | 1101 | 0 | 1 | 0 | 0 | 1 | 0 | 1 | 1 |
| 0110 | 1 | 0 | 0 | 0 | 0 | 1 | 1 | 1 | 1110 | 0 | 1 | 0 | 0 | 1 | 0 | 1 | 1 |
| 0111 | 0 | 0 | 1 | 0 | 1 | 1 | 0 | 1 | 1111 | 0 | 0 | 0 | 1 | 1 | 1 | 1 | 0 |

**Theorem 8** *Each function from set S4 can be used as a base for constructing composite Boolean function f and this function will have exact quantum query algorithm Q with complexity* $Q_E(A) = \frac{D(f)}{2}$.

**Proof.** Similar to the proof of Theorem 5.  □

### 3.2 Methods for Construction Low-Degree Polynomials

We are going to solve the problem of creation low-degree Boolean function from the other side - by construction of a low degree polynomial with Boolean values, and then define Boolean function by matching each possible input with corresponding value of the representing polynomial.

## Construction of a Polynomial of Degree 2 and Non-Boolean Range of Values

Let us have 9 variables divided into 3 groups by 3 variables in each. Appropriate graphical interpretation is described as follows (see the Figure 4):

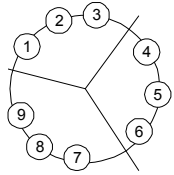 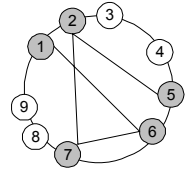

**Fig.4.** Division of 9 variables in 3 groups     **Fig.5.** Connections of variables

- each variable is presented as a point
- variable with the value 1 is depicted with colored point, 0 – with a transparent one
- there are exactly 3 groups of points: first group contains points 1,2,3, second – 4,5,6, third – 7,8,9.

Positioning of colored points obeys the next rules:
1. Points from different groups can be connected; points of the same group are never connected with each other.
2. A point from one group is allowed to be connected with the only point from another group, for example, if $x_1$, $x_2$, $x_5$, $x_6$, $x_7$ are „1" and others are „0", then a legal positioning of points is shown in the Figure 5. Here pairs (i,j) and (j,i) are identical and describe the same pair of points.

We are interested in the number of different pairs allowed to stand together. Given a set of indices of 1-valued variables, one way to find their connections is:

let the number of variables, $n=n_1+n_2+n_3$, be divided in 3 groups with $n_1$, $n_2$ and $n_3$ variables in each. Order them in descending order $n_1 \geq n_2 \geq n_3$. We are given a polynomial:

$$p(x_1,\ldots,x_n) = \sum_{i \in [n]} x_i - \sum_{\substack{i,j: \\ i \neq j, \\ (i,j) \in S}} x_i x_j \qquad (1)$$

where $\sum_{i \in [n]} x_i = |x|$ and $\sum_{\substack{i,j: \\ i \neq j, \\ (i,j) \in S}} x_i x_j$ is number of pairs, where $x_i = x_j = 1$ and (i,j) belongs to S- a set of pairs of connected points. In other words, $\sum_{\substack{i,j: \\ i \neq j, \\ (i,j) \in S}} x_i x_j = |S|$. Let us take an arbitrary input x, such that $|x| = k = k_1+k_2+k_3$, where $k_1$ is number of „1" in one group, $k_2$ and $k_3$ in other two, $k_1 \geq k_2 \geq k_3$.

Take the smallest of three, $k_3$, it is the number of connections corresponding group has with each of two other, $2*k_3$ in all. Then, a group with $k_2$ points has $k_2$ connections with the largest group. Thus, number of connections is $\sum_{\substack{i,j:\\i\neq j,\\(i,j)\in S}} x_i x_j = |S| = 2k_3 + k_2$. $\sum_{i\in[n]} x_i = k_1 + k_2 + k_3 = |x|$. Polynomial p(x) value is restricted from both left and right as follows from:

**Lemma 1.** $0 \leq p(x_1,...,x_n) \leq n_1$, where $n=n_1+n_2+n_3$ and $n_1=max(n_1,n_2,n_3)$.

**Proof.**
$p(x) = (k_1 + k_2 + k_3) - (2k_3 + k_2) = k_1 - k_3$
$0 \leq k_1 - k_3 \leq k_1 \leq n_1$

□

Returning to 9-variable polynomial
$p(x_1,x_2,x_3,x_4,x_5,x_6,x_7,x_8,x_9) = \sum_{i\in[9]} x_i - \sum_{\substack{i,j:\\i\neq j,\\(i,j)\in S}} x_i x_j$, that we have already mentioned, has a range of values {0,1,2,3}, as follows from Lemma 1.

The next task is to find an appropriate polynomial $p_b$ of degree 2 that would transform the set {0,1,2,3} to {0,1}, for example $p_b(z) = \frac{1}{2}z^2 - \frac{3}{2}z + 1$, $p_b(0) = p_b(3) = 1$,
$p_b(1) = p_b(2) = 0$. Hence, combined polynomial $p_b(p(x))$ has Boolean values and the degree $deg(p_b(p(x))) = 4$.

Define a Boolean function $f_9(x_1,...,x_9) = p_b(p(x_1,...,x_9))$, $D(f_9) = 9$, this fact is provided by sensitivity on zero input $|x|=0$, $f_9(x)=1$. Flipping any zero to 1 will change function's value to 0. Hence, $deg(f_9)=4$, $D(f_9)=9$.

**Generalization of the Approach**

Let us generalize the idea described earlier. Enlarge the number of variables, but still divide them in 3 groups. It appears that best results come from the case of N = n+n+n variable functions (n variables in each group). For each input x we are given (easy to determine) a set of pairs S describing connection of points. Define the polynomial

$$p(x_1,...,x_N) = \sum_{i\in[N]} x_i - \sum_{\substack{i,j:\\i\neq j,\\(i,j)\in S}} x_i x_j.$$

As follows from Lemma 1, $0 \leq p(x) \leq n$.

To show another example of low-degree function, let us take polynomial $p(x_1,...,x_{21})$ with the range of values {0,1,2,3,4,5,6,7}. There is a polynomial of degree 6 that we will use to transfer a set {0,1,2,3,4,5,6,7} to {0,1}, for example, $p_b(x) = -\frac{1}{144}x^6 + \frac{7}{48}x^5 - 1\frac{19}{144}x^4 + 3\frac{15}{16}x^3 - 5\frac{31}{36}x^2 + 2\frac{11}{12}x$. In general, for any odd k there is a polynomial of degree (k-1) that transforms $\{0,...,k\}$ to $\{0,1\}$. $p_b(p(x))$ values are from the range $\{0,1\}$, for corresponding $f_{21}(x) = p_b(p(x))$: $\deg(f_{21})=12$, $D(f_{21})=21$.

Using similar scheme, define Boolean function $f_{15}$, such that $D(f_{15})=15$, $\deg(f_{15})=2*4=8$ and $f_{45}$ such that $D(f_{45})=45$, $\deg(f_{45})=2*14=28$. A general form of this method is formulated as

**Lemma 2**. *For each odd k>1 there exists 3k-variable Boolean function f with D(f)=3k and deg(f)=2(k-1).*

**Tripple Function Method**

*Example 1*. 12-variable function $f_{12}(x_1,..., x_{12})$ with $D(f_{12})=12$ and $\deg(f_{12})=6$.
First, define a 4-variable polynomial of degree 3:
$p_4(x_1,x_2,x_3,x_4) = (x_1x_2 + x_2x_3 + x_3x_4 + x_1x_4) - (x_1x_2x_3 + x_1x_2x_4 + x_1x_3x_4 + x_2x_3x_4)$ $p_4(x)$
range of values is $\{0,1\}$. Next, define a polynomial $p_{12}(x_1,...,x_{12}) = p_4(x_1,x_2,x_3,x_4) + p_4(x_5,x_6,x_7,x_8) + p_4(x_9,x_{10},x_{11},x_{12})$ with values from the range $\{0,1,2,3\}$. Last step is to choose an appropriate polynomial of degree 2 to transform a set $\{0,1,2,3\}$ to $\{0,1\}$, for example, $S(z) = \frac{1}{2}z^2 - \frac{3}{2}z + 1$, $\deg(S(p_{12}(x))) = 6$. Define corresponding Boolean function $f_{12}(x) = S(p_{12}(x))$:
$D(f_{12}) = 12$ as its block sensitivity for the input x, $|x|=12$ is 12. A function of the similar form has been described in [8].

*Example 2*. We will try to extend an idea of function from the previous example to the case of n variables. Take $n = 3r$ variables polynomial (divided in 3 groups of $r$ variables each)
$P_v(x_1,...,x_r,x_{r+1},...,x_{2r},x_{2r+1},...,x_{3r}) = P_a(x_1,...,x_r) + P_a(x_{r+1},...,x_{2r}) + P_a(x_{2r+1},...,x_{3r})$.

**Lemma 3.** *For each odd k>1 and for each t>1 there exists $3^{t+1} \cdot k$ -variable Boolean function f, $D(f) = 3^{t+1} \cdot k$ and $\deg(f) = 2^{t+1} \cdot (k-1)$.*

**Proof.**
**Step 1**. Choose an odd *k* and define a polynomial according to Lemma 2 in the form $\{0,1\}^{3k} \xrightarrow{2} \{0,...,k\} \xrightarrow{k-1} \{0,1\}$, degree of it equals to *2(k-1)*, number of variables is *3k*.

**Step 2**. Define *3r*-variables polynomial (variables are divided in 3 groups of *r* variables each):

$P(x_1,\ldots,x_r,x_{r+1},\ldots,x_{2r},x_{2r+1},\ldots,x_{3r}) = P_0(x_1,\ldots,x_r) + P_0(x_{r+1},\ldots,x_{2r}) + P_0(x_{2r+1},\ldots,x_{3r}),$

where $P_0$ has Boolean values and the degree $d$. P has a range of values {0,1,2,3} that can be transformed to {0,1} by appropriate polynomial of degree 2, thus deg(P)=2d. This kind of polynomial can be iterated, as a result we get $\deg(P) = 2^t \cdot d$ and the number of variables $3^t \cdot r$ for any t>0.

**Step 3.** Take r=3k and d=2(k-1). Boolean function f(x) represented by P(x) has $D(f) = 3^t \cdot 3k = 3^{(t+1)} k$ and

deg(f) = $2^t$ *2(k-1)= $2^{(t+1)}$ (k-1).

$$\forall k > 1, (k \bmod 2) = 1, \forall t > 1 : \frac{D(P)}{\deg(P)} = \frac{3^t k}{2^t (k-1)}$$

## 4  Conclusion

In this work we have shown advantageous exact quantum algorithms and functions represented by a polynomial with a low degree that promise to be useful for construction of quantum algorithms. All work we have accomplished confirms a thought of perspective direction. The aim of our future work is using described functions to create quantum query algorithms with advantages over classical counterparts.